\documentclass{llncs}

\usepackage[scaled=.95]{helvet}
\usepackage{graphicx}
\usepackage{color}
\usepackage{epsf}
\usepackage{epsfig}
\usepackage{colortbl}
\usepackage{url}

\begin{document}
\title{Accessibility in Information Retrieval}
\titlerunning{}  
%
\author{Leif Azzopardi\inst{1} \and Vishwa Vinay\inst{2}}
\authorrunning{Azzopardi and Vishwa}   
%
\tocauthor{Leif Azzopardi and  Vishwa Vinay}
\institute{Dept. of Computing Science,\\ University of Glasgow, \\Glasgow UK,\\
\email{leif@dcs.gla.ac.uk},\\
\and
Microsoft Research Cambridge\\
Cambridge, UK\\
\email{vvinay@microsoft.com}}

\date{}

\maketitle

\begin{abstract}
 This paper introduces the concept of accessibility from the field of transportation planning and adopts it within the context of Information Retrieval (IR). An analogy is drawn between the fields, which motivates the development of document accessibility measures for IR systems. Considering the accessibility of documents within a collection given an IR System provides a different perspective on the analysis and evaluation of such systems which could be used to inform the design, tuning and management of current and future IR systems.
\end{abstract}


\section{Introduction}\label{sec_introduction}
Information Retrieval is the area that deals with the storage, organization, management and retrieval of information, where the goal of continual research in the field is to find {\it better} methods of doing the same. In pursuit of this betterment evaluation has been instrumental in the development of IR Systems. While evaluation has typically focused on the effectiveness~\cite{vanrijsbergen79ir}, or the efficiency~\cite{witten99gigabytes} of the IR system, these are only two ways in which to assess the quality of an IR system. In this paper, we introduce a complementary view to evaluation which provides a higher level view of the IR system by focusing on the {\it accessibility} an IR system provides to the documents in a collection.

Accessibility is an abstract concept coined almost 50 years ago in the land use and transportation planning field ~\cite{hansen59access}, where it was defined as a measure of potential opportunities for interaction with resources like employment, schooling, shopping, dining, etc. Measuring the accessibility in this context enabled many studies (e.g. ~\cite{hansen59access}\cite{neuburger71access}\cite{wachs73access}\cite{dong06activitybasedaccess}) to be performed which examined, for example, how changes in the levels of accessibility to such opportunities affected the urban area (in terms of economic impact, social changes and so forth). The results of such studies provide valuable information to transportation planners and city designers in the development of land use and transportation systems. Before this, planners and designers would focus on measures which were based on the effectiveness and efficiency of the transportation system (such as, the travel time between particular locations). However, accessibility provided a different perspective, and while related to effectiveness and efficiency, it takes a more abstracted or high level view on the evaluation of transportation systems, considering more general concerns relating to access, instead of focusing on specific instances.

Their definition of accessibility\footnote{Accessibility is also a key concept in other areas but defined differently. For instance, the disability rights movement advocates equal access to social, political, and economic life which includes not only physical access but access to the same tools, services, organizations and facilities. Another example is the World Wide Web Consortium (W3C)'s Web Accessibility Initiative (WAI), which is aimed to improve the accessibility of the World Wide Web for people using a wide range of user agent devices, not just standard web browsers. However, accessibility in these contexts concentrates on the physical aspects of accessing the information, and even extends to  issues regarding usability and mobility.} considers the accessibility of opportunities at locations in a physical space (such as a city). The transportation system is the means by which opportunities are made accessible (i.e., the road network and the bus, cycle path and a bicycle, etc). In this context, the main consideration in the design and management of the transportation system is to look beyond efficiency and effectiveness and to consider the accessibility of opportunities given a certain distance or the generalized cost the user is willing to incur to reach these opportunities and the desirability of these opportunities.

In the context of Information Retrieval, an analogy of accessibility can be made as follows. Instead of an actual physical space, in IR, we are predominately concerned with accessing information within a collection of documents (i.e., information space), and instead of a transportation system, we have an Information Access System (i.e., a means by which we can access the information in the collection, like a query mechanism, a browsing mechanism, etc). The accessibility of a document is indicative of the likelihood or opportunity of it being retrieved by the user in this information space given such a mechanism. For example, in a hyper-linked collection exposed by a browsing-based system, a page with no incoming or outgoing links will have no accessibility. Conversely, a page with thousands of incoming links would be very accessible. Here, we consider the accessibility of documents given an IR system, where documents are accessed by querying the system. Each query provides a different ordering in which to access the documents in the information space. Much like a particular bus taking a pre-defined route through a city. However, unlike in the physical space, in the information space, there is no constraint imposed by the user's current location (i.e., at a particular document) because the IR system facilitates access to the collection regardless of location. The IR system is like being at a bus stop where every possible bus route is available, (i.e., the universe of all possible queries), and we can select any route desired, at any time. While this makes every document potentially accessible, the choice of route and distance the user is willing to travel will affect just how accessible documents are in the information space.

In this paper, our main contribution is the introduction of the concept of accessibility and the proposal of how to measure accessibility in this context. To do so, we first describe the related research in Section ~\ref{sec_related_work} and draw upon the extensive body of work in transportation planning and land use to provide the basis in developing measures of accessibility for IR system. Then in Section ~\ref{sec_measures}, we propose two IR based accessibility measures that are analogous to those in the field of transportation. The introduction of accessibility presents many different possibilities and challenges which can not be fully addressed here, so we summarize this initial contribution in Section ~\ref{sec_conclusion}.

\section{Related Work}\label{sec_related_work}
In Hansen's seminal paper~\cite{hansen59access} on measuring accessibility in transportation planning and land use, he defines how accessibility could be measured:
\begin{small}
\begin{quotation}a measurement of the spatial distribution of activities about a point adjusted for the ability and the desire of people or firms to overcome spatial separation. More specifically, the formulation states that the accessibility at point 1 to a particular type of activity at area 2 (say employment) is directly proportional to the size of the activity at area 2 (number of jobs) and inversely proportional to some function of the distance separating point 1 from area 2. The total accessibility to employment at point 1 is the summation of the accessibility to each of the individual areas around point 1. Therefore, as more and more jobs are located nearer to point 1, the accessibility to employment at point 1 will increase.\end{quotation}
\end{small}

Key to this definition is the notion that as opportunities become further away the less accessible they become, and that by considering all possibilities to opportunities subject to the cost function based on the distance apart, provides a measure of accessibility. Essentially, this measure quantifies the \emph{potential of opportunities for interaction}~\cite{hansen59access}. In the context of IR, the opportunities are the documents in the information space, and we wish to capture the \emph{potential of documents for retrieval}.

\subsection{Measures of Accessibility in Transportation Planning}
There are numerous measures of accessibility that have been proposed in the field of Transportation Planning; the simplest and most popular measures are the Cumulative Opportunity Measures and Gravity Based Measures.

\emph{Cumulative Opportunity Measures} also known as Isochrone measures count the number of opportunities that can be reached within a given travel time, distance, or generalized cost~\cite{wachs73access}. An example application of the measure is ``the total number of dining opportunities within 400 metres''. The advantage of this measure is that it is intuitive and easy to compute. However, the measure is sensitive to the size of the range (around the point of interest) to be considered, and the representation of the opportunities.

First derived by~\cite{hansen59access}, \emph{Gravity Based Measures} provide a general method for measuring accessibility, which is widely used. They differ from cumulative based measures in that they include a cost function within the calculation. Generally, the cost function takes the form of a negative exponential function (as described by \cite{hansen59access}, above), such that opportunities that are further away will have a lower impact on the final accessibility value. By ``further'', it is meant in terms of time, distance or generalized cost.

While more sophisticated measures have been developed, such as \emph{Utility Based Measures}~\cite{neuburger71access} and \emph{Activity Based Measures}~\cite{dong06activitybasedaccess}, we shall only be considering the former two methods in this work as they are the most widely used and accepted measures in transportation and planning. Thus, it seems reasonable to use these as a starting point to determine if they can be useful and informative in IR, before developing more sophisticated measures.

\subsection{Accessibility in Information Retrieval}
Accessibility issues in IR have focused on restrictions (physical and virtual) to index or retrieve information, whether this is because of a physical impairment~\cite{fajardo06deafaccess}, restricted access due to security clearance~\cite{hawking04challenges}, or the inability to crawl portions of the web~\cite{bailey00access}. In each case, documents are inaccessible to the user or the system because of some physical or virtual limitation. For instance, in the latter case, the inability to crawl a web site means that certain documents are not indexed by the IR system, and therefore are not accessible to the user via the IR system. Recently, it was posited that the ``searchability'' of a web site would be affected by how easily pages can be crawled and how well the search engine matches and ranks them~\cite{upstill02searchability}. Searchability and accessibility are therefore very similar concepts. However, we are concerned with the influence of the IR system on accessing documents.
Others (e.g.~\cite{garcia04access}\cite{buttcher06access}) have considered how documents are accessed from the index in the retrieval process to facilitate more efficient retrieval by considering processor, disk and memory constraints. For instance ``access-ordered indices''~\cite{garcia04access} are where the documents which are more likely to be returned at higher ranks are placed before those that are not likely to be returned at higher ranks. Another example, is the caching of queries~\cite{bubba07querycaching}, in web search engines, where results pages are cached in response to popular queries in order to facilitate efficient access.

In essence, IR is all about \emph{accessing information}, and \emph{how the information is accessed}. Our work is focused on measuring the accessibility of documents in the collection given the IR system used to access these documents. This is different from past work, in that we are specifically examining the influence of the IR system to restrict or promote access to the information within the collection as opposed to other restrictions. This paper hopes to establish the idea of accessibility as an integral concept in the field by highlighting its potential in the practical task of developing, building, and optimizing IR Systems, as well as diagnostics and evaluation.


\section{Measuring Document Accessibility}
\label{sec_measures}
Given a collection $\bf D$, an IR system accepts a user query $\bf q$ and returns a ranking of documents $\bf R_q$, which are deemed to be relevant to $\bf q$ from within $\bf D$ by the IR system. We can consider the accessibility of a document as a system dependent factor that measures how retrievable it is, with respect to the collection $\bf D$ and the ranking function used by the IR system. Using the analogy of transportation, entering a query is like to choosing a particular bus, where the order of documents returned are like the order of destinations reached for that given bus route. Opportunities to interact with resources while traveling along the route are reflected by going through the documents returned in the ranking $\bf R_q$. The accessibility of the resources (i.e., documents) is dependant on the willingness of the user to travel a certain distance along the route (i.e., traverse down the ranked list) and all the queries that users are likely to travel along. So, by adapting the measures from transportation planning, we propose a general measure of the accessibility of a document, as:
\begin{equation}
\label{eq:the_equation}
A({\bf d}) = \sum_{{\bf q} \in {\bf Q}} o_q \cdot f(c_{dq}, \theta)
\end{equation}
\noindent where {$o_q$ denotes the likelihood of expressing query $\bf q$ from the universe of queries $\bf Q$ and $f(c_{dq}, \theta)$ is a generalized utility/cost function where $c_{dq}$ is the distance associated with accessing $\bf d$ through $\bf q$ which is defined by the rank of the document, and $\theta$ is a parameter or set of parameters given the specific type of measure. 

A cumulative based measure can then be defined as follows: $\theta = c $, where $c$ denotes the maximum rank that a user is willing to proceed down the ranked list. The function $f(c_{dq}, c )$ returns a value of $1$ if $c_{dq} \leq c$ (with the top-most position considered as rank $1$), and $0$ otherwise. So, if returning a document in response to a given query has a distance greater than $c$ associated with it, then it is considered unaccessible (for this query). For another query however, the document may be accessible because the cost of accessing it is within the distance $c$. Alternatively, the document could be considered accessible for the same query but to a user who has a higher cost threshold. Since all the documents within the cutoff defined by $c$ are equally weighted, this type of measure emphasizes the number of times the document can be retrieved within that cutoff over the set $\bf Q$.


A gravity based measure can also be defined by setting the function to reflect the effort of going further down the ranked list, such that the further down the ranking the less accessible a document becomes. There are numerous ways in which such a function could be determined. Here, we adopt the function suggested in \cite{hansen59access}, where the accessibility of the document is inversely proportional to the rank of the document, such that:
\begin{equation}
f(c_{dq},  \beta) = \frac{1}{(c_{dq})^{\beta}}
\end{equation}
\noindent where, the set of parameters $\theta$ includes $\beta$ which is a dampening factor that adjusts how accessible the document is in the ranking. Interestingly, if the $\beta$ parameter is set to one, then accessibility of the document for the given query is equivalent to the reciprocal rank of the document, which is related to the (expected) search length \cite{cooper68esl}. When there is only one relevant document, the expected search length is equivalent to the reciprocal rank of the document. Intuitively, the expected search length (ESL) and accessibility of documents is related, because the expected search length corresponds to how many irrelevant documents have to be examined in order to find the relevant documents. The expected search length to a particular document is proportional to the accessibility of the document for a given query. However, what the accessibility measure captures is more general, i.e., how retrievable the document is given all possible/likely queries regardless of relevancy, but this link to ESL and reciprocal rank appears to provide a connection between accessibility and effectiveness. As we have previously mentioned this direction is left for future work.

Given either measure, $A({\bf d})$ provides an indication of the opportunity of retrieving $\bf d$. This value can be obtained for each document ${\bf d} \in {\bf D}$ so that we can compare whether there is more opportunity to retrieve one document over another. Using this measure to compare groups of documents has potential to aid in the design, management and tuning of retrieval systems in a number of ways. Imagine that for a given collection of documents and a given IR system, the average $A({\bf d})$ of a set of documents is extremely high, while for another set of documents the average $A({\bf d})$ is very low. Perhaps, the first set of documents was a group of site entry pages, and our system has a prior towards such pages, thus we would expect these pages to have a higher $A({\bf d})$. In this case, it is desirable that these documents are so accessible. On the other hand, if the set of highly accessible pages was composed of spam pages, because these pages have used ``tricks'' to artificially inflate the number of queries for which they are retrieved, then this is not desirable and the system needs to be adjusted. Alternatively, if there is a set of documents which are virtually inaccessible in the collection, then it is a management decision to decide whether these documents should be included in the index or not.

At a higher level, the measure $A({\bf d})$ motivates questions regarding how accessible documents in the collection should be, and whether we are interested in trying to ``hide'' or ``promote'' certain documents within the collection. Or whether we should adopt an approach that ensures access to the information is free from bias, i.e. ``universal access''\footnote{As previously mentioned, the disability rights movement advocates equal access  and terms this notion as universal access.} so that \emph{any document is as accessible as any other document} in the collection. This provides a novel framework for measuring document accessibility, which enables the consideration of such questions and issues.

\section{Conclusion and Future Work}\label{sec_conclusion}
The main contribution of this paper is the introduction of the concept of accessibility and quantifying the accessibility of documents in the collection given a particular IR system. Measures of accessibility are not performance measures like effectiveness or efficiency, but instead are measures of the \emph{potential of documents for retrieval}  (a.k.a., their retrievability~\cite{azzopardi2008retrievability}). This abstraction provides a novel way to quantify and detect different levels of accessibility within the collection imposed by the IR system. For a system administrator, this could prove to be very useful in designing, managing and tuning the IR system  (see \cite{azzopardi2008doc_access} for empirical examples).

This work represents the initial step towards formalizing accessibility and developing accessibility measures for information spaces, in IR and more generally for any Information Access system. However, there are many open problems, challenges and issues which have arisen as a result of this work. Further research needs to be conducted in two main directions: \begin{enumerate} \item the calibration, computation and estimation of document accessibility measures, and \item the application of document accessibility measures. \end{enumerate}

\paragraph*{Acknowledgements} The authors would like to thank Stephen Robertson, Keith van Rijsbergen and Murat Yakici for their helpful and insightful comments and suggestions.

\bibliographystyle{abbrv}
\bibliography{refs}
\end{document}